\documentclass[apj,iop]{emulateapj}

\pdfoutput=1

\usepackage{amsmath}
\usepackage{amssymb}
\usepackage{latexsym}
\usepackage{wasysym}
\usepackage{url}

\shorttitle{ORIGAMI}
\shortauthors{Falck, Neyrinck, \& Szalay}

\begin{document}

\newcommand{\figscl}{0.34}
\newcommand{\org}{{\scshape origami}}
\newcommand{\fof}{{\scshape fof}}
\newcommand{\noned}{$N_{\rm 1D}$}
\newcommand{\ihmpc}{\,$h$\,Mpc$^{-1}$}
\newcommand{\hmpc}{\,$h^{-1}$\,Mpc}

\title{ORIGAMI: Delineating Halos using Phase-Space Folds}
\author{Bridget L. Falck, Mark C. Neyrinck, and Alexander S. Szalay}
\affil{Department of Physics and Astronomy, Johns Hopkins University, 3400 N Charles St, Baltimore, MD 21218}

\begin{abstract}
We present the \org\ method of identifying structures, particularly halos, in cosmological $N$-body simulations. Structure formation can be thought of as the folding of an initially flat three-dimensional manifold in six-dimensional phase space. \org\ finds the outer folds that delineate these structures. Halo particles are identified as those that have undergone shell-crossing along 3 orthogonal axes, providing a dynamical definition of halo regions that is independent of density. \org\ also identifies other morphological structures: particles that have undergone shell-crossing along 2, 1, or 0 orthogonal axes correspond to filaments, walls, and voids respectively. We compare this method to a standard Friends-of-Friends halo-finding algorithm and find that \org\ halos are somewhat larger, more diffuse, and less spherical, though the global properties of \org\ halos are in good agreement with other modern halo-finding algorithms.
\end{abstract}

\keywords{dark matter -- galaxies: halos -- large-scale structure of Universe -- methods: numerical}

\section{INTRODUCTION}

Cosmological $N$-body simulations allow one to calculate the time-evolution of an initial density field, discretized as a set of dark matter point particles with a given mass, from the distant past when the field was quite smooth to the present-day hierarchy of halos, filaments, walls, and voids. Identifying these structures remains one of the key challenges to the process of comparing these simulations to observations of galaxies and clusters. Though some efforts have been made to identify complex structures such as filaments and walls~\citep{ara07,hah07,for09,bon10,sha11JCAP}, most of the focus is on identifying the dark matter halos in which galaxies reside or the voids that comprise most of the cosmological volume.

The two best-known ways of identifying halos in simulations are the Spherical Overdensity~\citep[SO,][]{pre74,lac94} and Friends of Friends~\citep[FOF,][]{dav85} methods. Since these were developed, the number of halo-finding algorithms has grown quite large~\citep[for an extensive listing, see][and references therein]{kne11}, but many of these modern methods rely at their core on the SO~\citep[e.g.,][]{kly97,kno09,pla10,sut10} or FOF~\citep[e.g.,][]{got99,spr01,gar07,hab09,ras10} algorithms. Other methods work in phase-space~\citep{die06,mac09,beh11,ela11}, or in some other way group particles around density peaks~\citep{eis98,sta01,aub04,ney05,twe09}. In detail, these methods differ in terms of how densities are calculated, whether they perform any post-processing or ``unbinding'' procedures, whether they identify sub-structures, and implementation details such as code parallelization. Recently the ``Haloes Gone MAD'' comparison project has tested many of these methods on an equal footing, finding that the differences for the basic halo properties in a cosmological simulation are well within the expected error~\citep{kne11}.

We would like to note, however, that the agreement between different halo-finding methods for only the most massive halos, or for masses defined within some radius, is perhaps not unexpected, since one of the largest sources of variation between methods is the definition of the halo boundary or outer-edge. Often this is because the halo boundary depends quite strongly on the value of a free parameter in the algorithm, such as a density cut-off or (in the case of \fof) a linking length that effectively serves as a proxy for density~\citep{eis98,ney05,kne11,and11}. \citet{kne11} did not explicitly test the agreement among halo finders of halo boundaries, for which we expect a criterion could be designed that would show surprisingly poor agreement, such as the maximum Cartesian halo size which we use below.
Going to full six-dimensional phase-space~\citep{die06,mac09,kne11,beh11} allows impressive identification of distinct halo subhalo cores (as well as streams), although even here the boundaries of halos and subhalos can be ambiguous.
As pointed out by \citet{ShandarinEtal2011}, knowledge of the initial and final conditions of position coordinates is equivalent to knowledge of the full six-dimensional phase space in the final conditions for a classical Hamiltonian system, a fact which we exploit.

In this paper, we present the \org\footnote{{\centerline{Order-ReversIng\ \ \ \ \ \ \ \ \ \ }
\centerline{Gravity, Apprehended}
\centerline{Mangling Indices}}} structure-finding algorithm which finds halos by testing whether particles have undergone shell-crossing.
We set halo boundaries at their outer caustic, i.e.\ at the outermost phase-space fold, which \citet{ZukinBertschinger2010} have found to correspond well in an analytical model to a conventional density-based concept of a virial radius. The formation of structures in the universe has long been linked to the formation of caustics as matter piles up and forms pancakes or sheets~\citep[][p. 95]{zel70,pee80}. These caustics mark out the boundaries of multi-stream regions, i.e., locations in physical space for which the velocity field is multi-valued. Particles that have entered multi-stream regions are said to have undergone shell-crossing, and their dynamics become quite complicated as they settle into a bound structure~\citep{KofmanEtal1990,KofmanEtal1992,vog08,whi09,sha11JCAP,val11,vog11}.

To provide some intuition as to how the \org\ method works, consider that though usually particles are thought of as simple blobs of mass, they
can also be thought of as vertices of an initially regular grid (which is often the case for the initial conditions of $N$-body simulations; however, \org\ currently does not work for ``glass-like'' initial conditions).  Gravity
distorts this grid, causing some of its cells to collapse and invert
when shell-crossing structures form.  In three-dimensional position
space, in such shell-crossing regions, multiple cells overlap at
the same position.

However, in six-dimensional phase space, these cells never cross, assuming a numerical-error-free simulation of collisionless dark matter. Instead, a three-dimensional manifold, or sheet, stretches and folds in six dimensions, forming familiar large-scale structures when the velocity coordinates are projected out.  Extrapolating to time zero, the grid is exactly regular in position coordinates and all velocities are zero, so initially this sheet is flat in phase space. Then in subsequent gravitational evolution, the sheet folds without intersecting itself in phase space.  If there were such an
intersection, then two dark-matter particles with different initial
coordinates would have the same phase-space coordinates at a later
time, a contradiction for a Hamiltonian dynamical system
\citep{LandauLifshitz}.  This picture has also recently been explored
by \citet{ShandarinEtal2011} and \citet{AbelEtal2011}.  They use
tetrahedral tessellations on the initial grid to identify shell
crossings. \citet{AbelEtal2011} use this to measure densities within
the phase-space sheet, allowing for example a particularly clean
visualization of the cosmic web.  Our framework, however, keeps track
of the number of axes along which particles have crossed, enabling
structures to be classified as voids, walls, filaments and halos.

\begin{figure*}[htb]
  \begin{center}
    \includegraphics[width=\hsize]{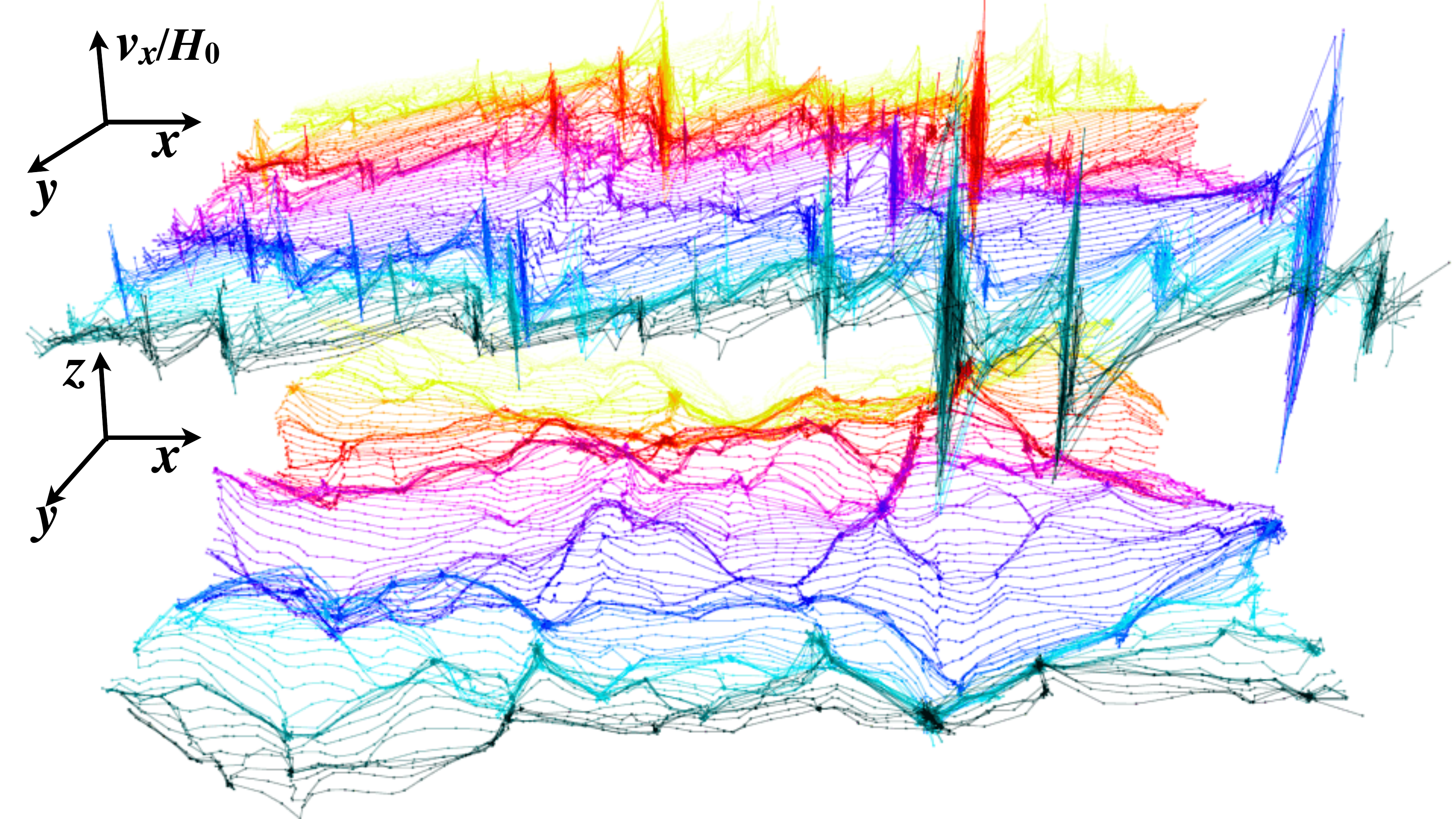} %scale=.4
  \end{center}  
  \caption{Distortion and folding in phase-space at the present epoch
    of a 128$^2$-particle sheet, roughly 100\hmpc\ on a side.  This
    is a quarter of a $256^2$-particle sheet, initially a
    two-dimensional flat slice through the 256$^3$-particle cubic
    lattice.  In the bottom sheet, particles are plotted at their
    familiar position coordinates, $(x,y,z)$; in the top sheet, the
    $z$ coordinate switches to $v_x/H_0$, the $x$-component of the
    velocity scaled with the Hubble constant. Rows of particles are colored according to their initial
    (Lagrangian) $y$-coordinate.}
  \label{fig:origami_wrapfig}
\end{figure*}

Figure \ref{fig:origami_wrapfig} illustrates the wrapping and
stretching that gravity imparts to an initially flat sheet of
particles in phase space.  We plot two projections of a Lagrangian
sheet of 256$^2$ particles from a full 3D 256$^3$-particle
cosmological simulation run to redshift zero. The bottom sheet
projects out the velocity coordinates and shows the familiar
$(x,y,z)$ coordinates.  The top sheet offers a peek into velocity
space, replacing the $z$ (vertical) coordinate with $v_x/H_0$, the $x$
(horizontal) component of the velocity scaled with the Hubble
constant.  Halos are visible as knots in the bottom sheet; in the top
sheet, they show up as furious spikes, which if zoomed into would
ideally exhibit a spiral structure
\citep[e.g., ][]{FillmoreGoldreich1984,Bertschinger1985,WidrowKaiser1993}.

Implicit in the name of our algorithm is an analogy to origami.  In
cosmological structure formation, as in origami, the ``sheet'' starts
out flat initially, and never crosses itself when viewed in phase
space.  But of course, the analogy is not complete.  Cosmological
sheets stretch, unlike origami sheets. Also, the dimensionality is
different: the folding of an actual two-dimensional origami sheet
occurs in three dimensions, whereas even in a two-dimensional
universe, folding of the cosmological sheet occurs in four
dimensions. 
If we flatten away the velocity coordinates, the situation is
analogous to ``flat origami,'' a restriction of origami in which the
end result of the folding is constrained to lie flat in a plane.  In
the cosmological case, dark-matter caustics correspond to creases in
the folded structure.  The field of flat origami has been studied
mathematically~\citep[e.g.\ ][]{Hull1994,Lang1996,Hull2002,Hull2006}, 
and in the discussion section we speculate on some applications of those results to large-scale structure.

We describe the \org\ method in Section~\ref{sec:method}, including how particles are tagged according to their morphology, how these particles are grouped into halos, and how halo properties are calculated. Our morphology classification results are given in Section~\ref{sec:morphology}. In Section~\ref{sec:fof} we compare the \org\ halo catalog to a standard \fof\ catalog, and in Section~\ref{sec:res} we study the effects of the mass resolution of the simulation. Finally, we discuss particular features and speculate on potential applications of \org\ in Section~\ref{sec:disc}, and we give concluding remarks in Section~\ref{sec:conc}.

\section{METHOD}
\label{sec:method}

Most fundamentally, the \org\ algorithm classifies particles according to their morphological structure into void, wall, filament, and halo particles, which we describe in Section~\ref{sec:tag}.  Usually, a catalog of structures is desired, which requires grouping these particles.  There are many ways to do this, and we explain our procedure, based on a Voronoi tessellation in Eulerian space, in Section~\ref{sec:group}. Finally, we describe our method of measuring halo properties in Section~\ref{sec:props}, which will be used to compare halo catalogs in the following sections.

\subsection{Morphology Classification}
\label{sec:tag}

\org\ classifies the morphologies of particles based on the number of orthogonal directions along which the Lagrangian ``origami'' phase-space sheet is folded.  Although the origami-folding occurs in full phase space, for the present paper we make use only of position space, taking advantage of the fact that the relative positions of particles within folded regions have been reversed with respect to the initial grid.  It may help for some purposes (such as demarcating subhalos) to include velocity information as well; however, as mentioned above, knowledge of the initial and final states of a classical Hamiltonian system is equivalent to knowledge of the final positions and velocities.  The initial state for a cosmological simulation run from an initial particle lattice is particularly simple and thus a natural choice of data to exploit.

A particle's \org\ morphology is determined by the number of axes along which particle-crossing has occurred. In one dimension, this is trivially tested by determining whether the final (Eulerian) positions of two particles are out of order with respect to their initial (Lagrangian) positions. 
This means that if we index the particles according to their
positions on the initial, regular Lagrangian lattice, particles $i$
and $j$ have crossed if $i<j$ but their order along that axis is
swapped, i.e.\ their Eulerian positions $x_i>x_j$. 
In two or three dimensions, we use this same criterion to detect
crossings along rows and columns of the initial Lagrangian lattice.
A particle $i$ has been crossed along the $x$ axis if there exists a
particle $j$ in its initial $x$-oriented row such that $(x_i - x_j)$
and $(i - j)$ have opposite signs, where again the indices $i$ and $j$
increase with initial $x$-coordinate. Note that we are taking
advantage of the lack of substantial vorticity on cosmological scales;
if a region in a simulation were able to rotate along three axes, our
algorithm would detect that as well.

Void, wall, filament, and halo particles are particles that have been
crossed along 0, 1, 2, and 3 orthogonal axes, respectively.  This number is a
particle's \org\ morphology index $M$.  The most natural set of axes
to use is the intrinsic Cartesian $x$, $y$, and $z$ axes of the
initial grid, but particle crossing may occur along other axes as
well.  In practice, using only the Cartesian axes seemed to detect
only about half of the particle crossings, in the sense that sets of
halo particles that should have been contiguous (see
Figure~\ref{fig:rhovsm} below) contained non-halo particles.  We
found that using three additional orthogonal triplets of axes filled these holes. Each triplet consists of one intrinsic $x$, $y$, or $z$ axis and two
$45\degr$-diagonal axes in the plane perpendicular to the intrinsic
axis.  The morphology index $M$ returned is the maximum $M$ among all
four sets of axes.  

\begin{figure}[htb]
\centering
\includegraphics[angle=90,width=\hsize]{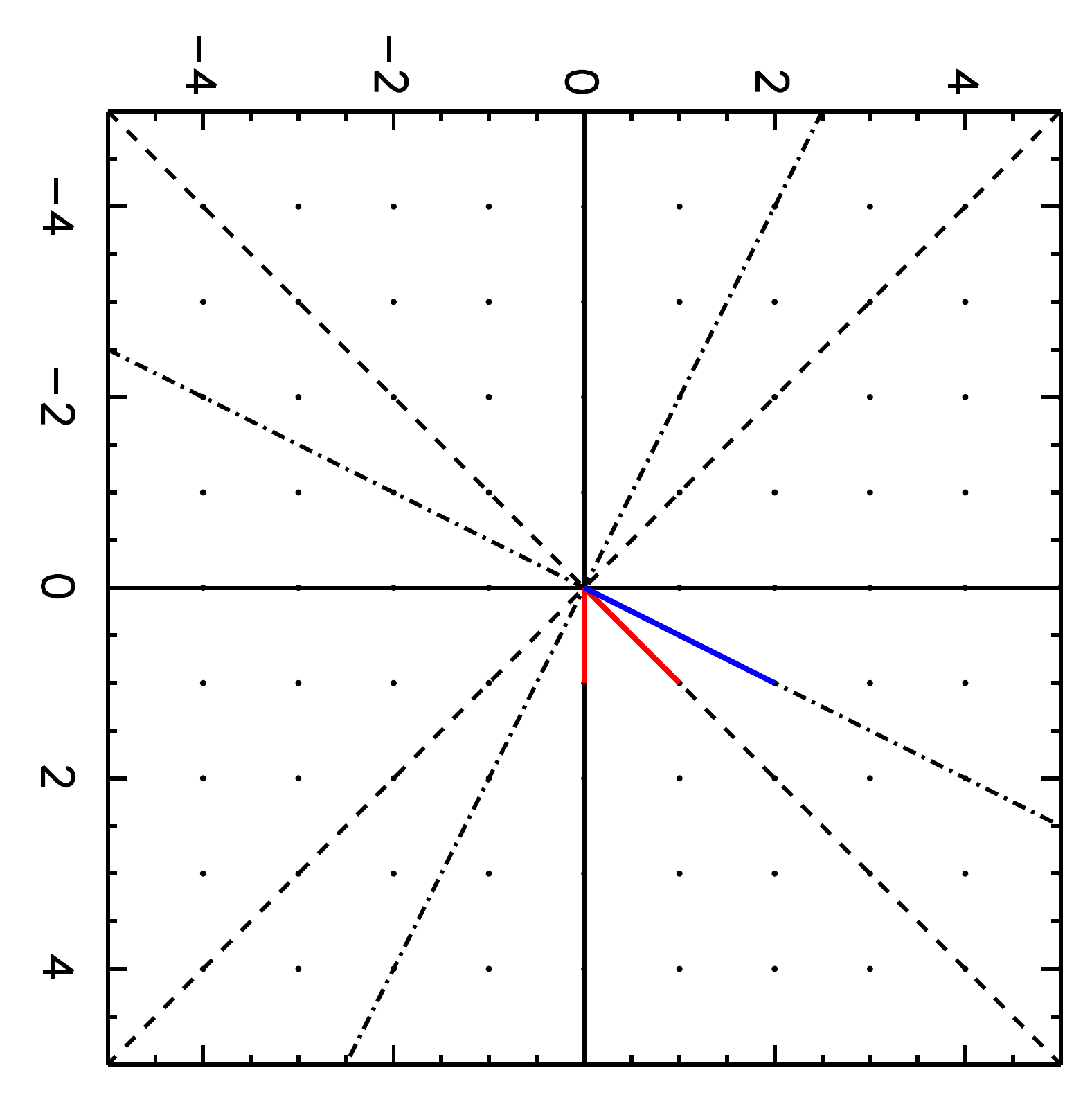}
\caption{Schematic of the initial particle separation along different sets of axes. The Cartesian axes \emph{(solid)} and a $45\degr$ diagonally-rotated set of axes \emph{(dashed)} have the smallest particle separations (shown in \emph{red}). A third set of axes \emph{(dot-dashed)} has the next-largest particle separation (shown in \emph{blue}). We use the Cartesian grid and the three sets of $45\degr$ rotated axes (where the rotation is in the $x$-$y$, $y$-$z$, and $x$-$z$ planes) to test for particle crossing. \label{fig:axes}}
\end{figure}

In principle, there is a huge number of ``higher-order'' axes along which particles might cross, which are at odd angles with respect to the Cartesian grid. As the ``order'' of a set of axes increases, so does its minimum initial particle separation (if a cubic initial grid is used). This increased particle separation decreases the likelihood that particles along these axes will interact and collapse into the same bound structure. Figure~\ref{fig:axes} shows a schematic of the initial particle separation along different choices of axes (here in two dimensions for clarity), including the Cartesian grid, axes rotated by $45\degr$, and a ``higher-order'' set of axes. Note that the initial particle separation is not the same along each axis for these rotated sets of axes, a property which is quite rare and leads to even larger initial separations.
 
We find that including particle-crossings detected along all 6 permutations of the extra set of axes shown in Figure~\ref{fig:axes} increases the number of halo particles by only 3\% or 4\% (depending on simulation resolution), and in all, only about 5\% of particles increase their morphology index $M$. These added particles have a negligible effect on the halo mass functions and a small effect on the halo sizes (defined in Section~\ref{sec:props}), growing the halos slightly. Though in principle many more higher-order axes could be used to find particle-crossings, these added particles do not change our conclusions while adding significant computational time to the morphology classification, so using only the 4 lowest-order sets of axes to determine \org\ morphology provides a good balance between strict completeness and computational efficiency.

The particle-crossing detection algorithm is simple and very fast.
For each particle, for all six axes ($x$, $y$, $z$, and the
$45\degr$-diagonal axes in the $x$-$y$, $y$-$z$, and $x$-$z$ planes), we
march forward (not backward, since all particles are tested) along the
initial lattice and test for any crossing along that axis.  Denoting
the number of particles in an initial row or column as \noned, we test
up to \noned/4 particles in each row, i.e.\ proceeding along 1/4 of the
initial lattice and stopping early if a crossing-detection occurs.  We do
not approach 1/2 of the lattice, since at that separation a false
detection could occur as an artifact of periodic boundary conditions (PBC).  Thus the number of distance tests required, in worst case (when a full
fourth of a row is tested for each particle) is $2\times6N_{\rm 1D}N/4 =
3N^{4/3}$, where the total number of particles in the simulation is $N=N_{\rm 1D}^3$.  The factor of 2 is from distance tests from PBC-corrections, and the factor of 6 is from the total number of axes tested.

\org\ morphology can be determined at any of the output redshifts of the simulation, but we only use the positions at that redshift to test for particle-crossing. This means that if particles have crossed at an earlier redshift, we do not use that information but instead look for new particle-crossings. We leave the study of redshift-dependent \org\ morphology, for example whether it will aid in substructure identification, to future work; however, we note that retaining the \org\ morphology of the next-to-last snapshot (i.e. not allowing $M$ to decrease) increases the number of $z=0$ halo particles by a few percent. In Section~\ref{sec:results} we present the morphology results and halo catalogs using $z=0$ only.

\subsection{Grouping Halo Particles}
\label{sec:group}

Once all particles have been given a morphology classification, we group the halo particles using a Voronoi/Delaunay tessellation, which provides a natural density estimate \citep[Voronoi Tessellation Field Estimator (VTFE),][]{sch00,van09} and set of neighbors for each particle. 
A Voronoi tessellation partitions space into cells, such that all points inside a particle's Voronoi cell are closer to that particle than to any other. The Delaunay tessellation is the dual of the Voronoi tessellation and divides a 3-dimensional volume into a set of tetrahedra (or a 2-dimensional area into triangles) that connect particles, such that the particles in two adjacent Voronoi cells are connected in the Delaunay tessellation. The VTFE density at each particle is given by $\delta_{\rm VTFE} = \bar{V}/V-1$, where $V$ is the particle's Voronoi cell volume and $\bar{V}$ is the average of $V$ among all particles. 

We group halo particles that are connected on the Delaunay tessellation, but to prevent long strings of halos from being linked, we first require that halos contain only one ``core'' or set of connected halo particles above some VTFE density threshold. This becomes the only parameter in the \org\ algorithm, which we set to 200 times the mean density. In this first round, any particles meeting this criterion that are Delaunay neighbors are given the same halo ID. We then add the lower-density halo particles that are connected to the halo cores by iteratively adding Delaunay neighbors until all connected particles are associated. Finally, we group any leftover halo particles that are not connected to a core but are connected to each other on the tessellation.

\begin{figure}
\centering
\includegraphics[angle=90,width=\hsize]{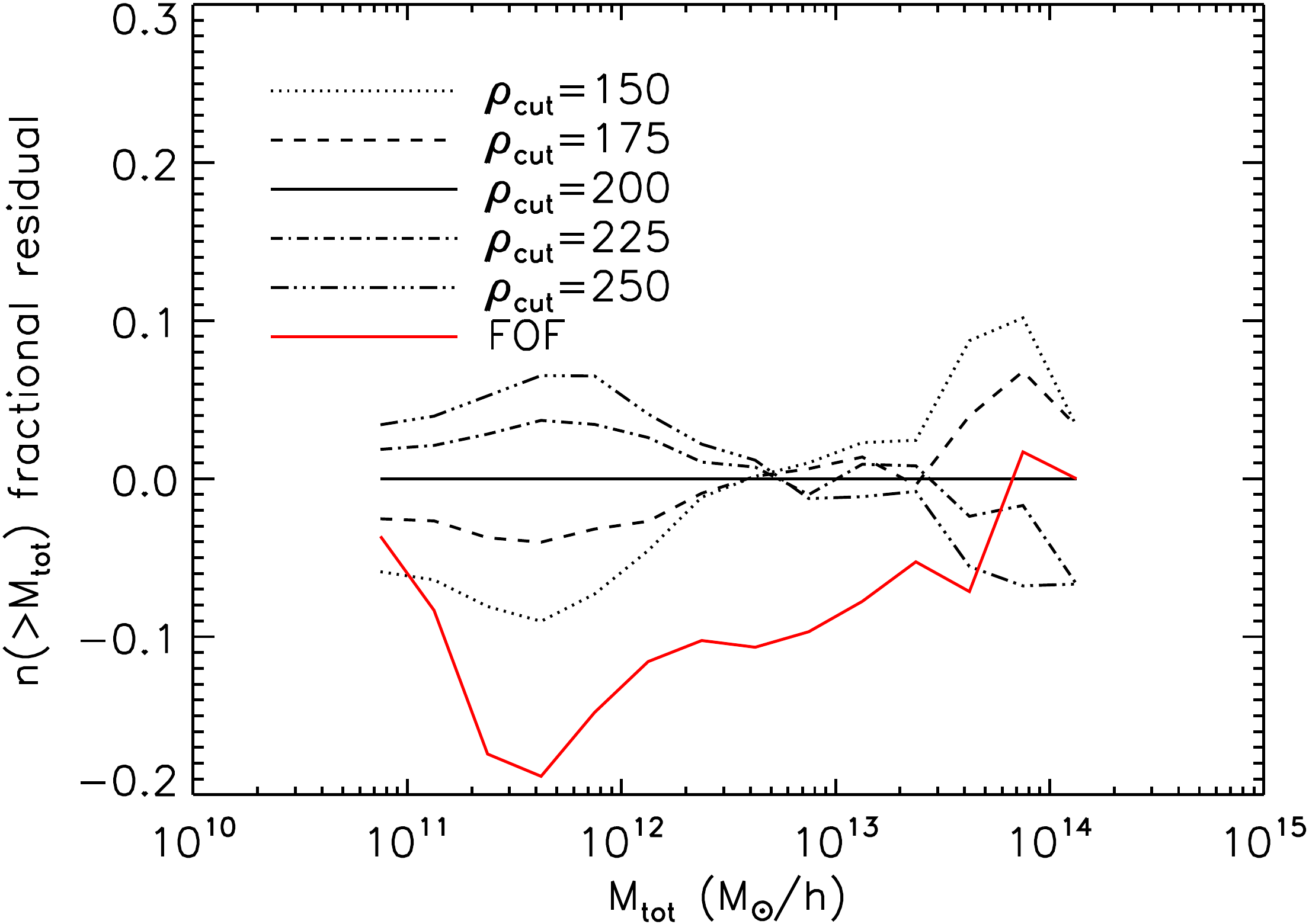}
\caption{The effect of different values for the ``core'' density threshold shown as fractional residuals from the cumulative mass function of the $\rho_{cut}=200$ value used throughout the paper. Relaxing this value produces fewer small halos, and increasing it produces more small halos, though the difference is small.\label{fig:mtot_core}}
\end{figure}

We find that the core density threshold has a small effect on the mass functions of the grouped halos and has negligible effect on the distribution of halo sizes (defined in Section~\ref{sec:props}). The effect on the mass function is shown in Figure~\ref{fig:mtot_core}. A smaller threshold produces fewer small halos and a higher threshold leads to more small halos, as might be expected. The differences in the cumulative mass functions for $\rho_{cut}=150$ and $\rho_{cut}=250$ are within the 10\% level compared to the value of $\rho_{cut}=200$ used throughout the paper, and they are smaller than the difference with \fof. See Section~\ref{sec:fof} for more discussion of \org\ vs. \fof\ mass functions.

The grouping procedure thus puts all of the tagged halo particles into individual halos. Since the tagging procedure establishes halo boundaries by identifying halo particles, no post-processing to remove unbound particles is performed. In principle, there is no lower limit to the number of particles in an \org\ halo, though it is hard to imagine isolated particles being classified as a halo as described above. However, we require that halos contain at least 20 particles in order to be included in our final catalog.

\subsection{Measuring Halo Properties}
\label{sec:props}

In this section we define the halo properties used to compare halo catalogs in Sections~\ref{sec:fof} and~\ref{sec:res}. These will be used to calculate the halo properties for both the \org\ and \fof\ halo catalogs. One of the most important is the definition of the halo center, since it affects many other calculated properties. We define the center to be the average position of the halo particles, weighted by their VTFE density. This greatly reduces the dependence of the location of the center (and many halo properties) on the low-density outer regions of non-spherical halos, compared to a non-weighted average.

We determine $R_{200}$, the radius beyond which the density drops below 200 times the critical density ($\rho_{crit}$), and $M_{200}$, the mass within this radius, by sorting the particles according to their radius from the halo center and determining the maximum radius for which the density within this radius is greater than $200\rho_{crit}$. 
If no radius meets this criterion for a specific halo, we consider $R_{200}$ and $M_{200}$ undefined and set their values to zero. We look at mass functions of $M_{200}$, which is a common definition of halo mass found in the literature \citep[see, e.g.,][]{kne11}.

Though $M_{200}$ is commonly used to define the mass of halos, we also look at the total mass of all halo particles in order to include the full extent of the halo, since one of the key differences of \org\ compared to other methods is its dynamical definition of the halo boundary. \citet{kne11} found that $v_{\rm max}$, the peak of the rotation curve, is a stabler definition of halo mass across different halo finders, but this is also due largely to its insensitivity to halo boundaries. We use the more discriminating total mass of a halo in our comparisons, since a major motivation for our algorithm is to create an objective definition of the halo boundary.

The final property we will compare is the halo size, which we define as the maximum diameter along the Cartesian $x$, $y$, and $z$ directions. This choice is preferred to other common measures such as $R_{200}$ because it captures the outer regions of the halo.
The halo size is calculated in the following way: first we transform the halo particle positions to a local coordinate system to account for PBC; we then calculate $s_x$ = max($x$) - min($x$) for the $x$, $y$, and $z$ coordinates; and finally we take the halo size to be the maximum of ($s_x$, $s_y$, $s_z$). We choose this definition over, for example, the maximum radius of the particles in the halo because it does not depend on the definition of the halo center.

\section{RESULTS}
\label{sec:results}

We first present the morphology classification that \org\ gives to particles in an $N$-body simulation in Section~\ref{sec:morphology}. 
We then focus on comparing \org\ to \fof\ in Section~\ref{sec:fof}, followed by an investigation of the effects of resolution on both \org\ and \fof\ catalogs in Section~\ref{sec:res}.

\subsection{Morphology}
\label{sec:morphology}

\begin{figure}
  \begin{center}
    \includegraphics[width=\hsize]{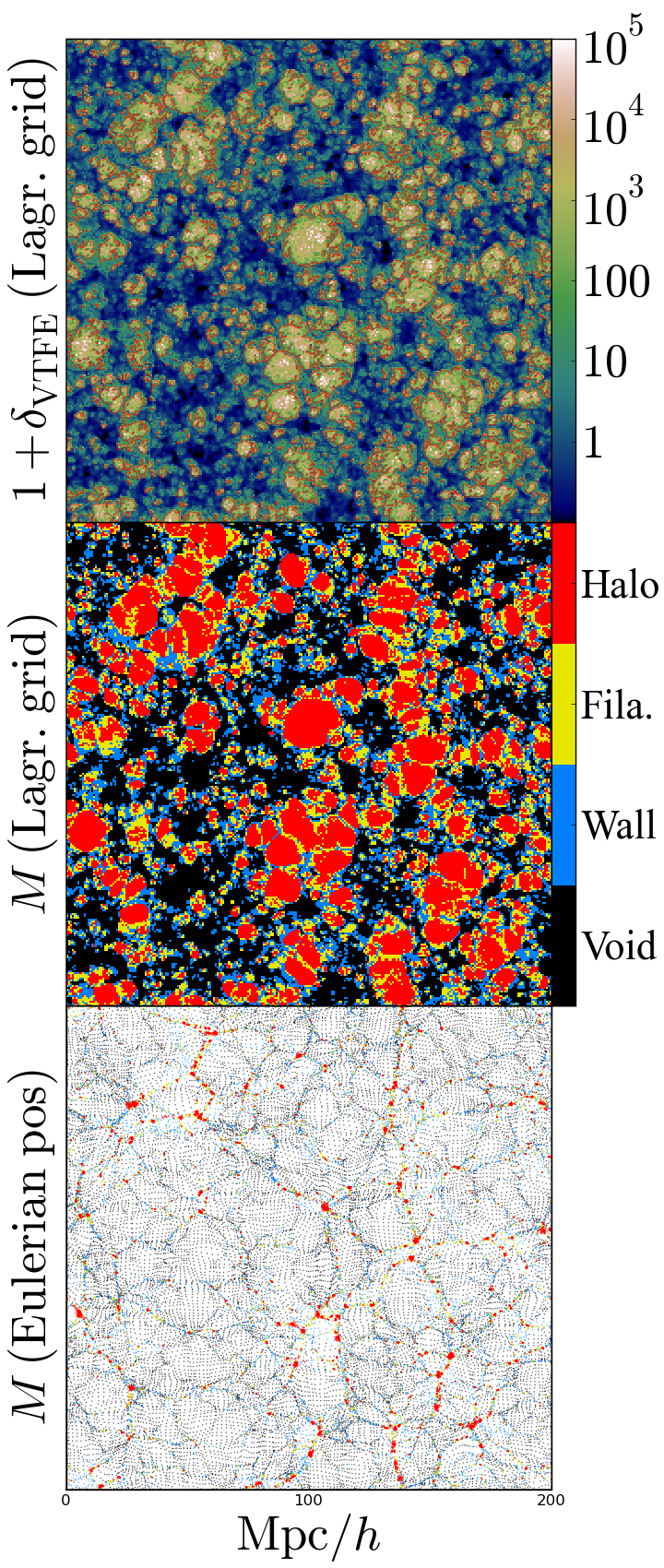} %.35
  \end{center}  
  \caption{Redshift-zero quantities measured from a 256$^2$ sheet of
    particles that share the same $z$-coordinate in the
    initial-conditions lattice. The top two panels are shown in
    Lagrangian coordinates, in which each particle is a pixel in a
    256$^2$ image. {\it Top}: Voronoi-tessellation density estimates,
    at each particle (see text for explanation). A red contour is
    drawn at $1+\delta=61$ (see text for details).  {\it Middle}:
    \org\ morphology indices $M$. {\it Bottom}: Here the $M$-colored
    particles (0-3 are shown in black, blue, yellow, and red) are
    plotted at their $(x,y)$ Eulerian coordinates.}
  \label{fig:rhovsm}
\end{figure}

In this section we use a 256$^3$-particle simulation with a box size of 200~$h^{-1}$Mpc and standard $\Lambda$CDM cosmology ($h=0.73$, $\Omega_M = 0.3$, $\Omega_\Lambda = 0.7$, $n_s = 1$, and $\sigma_8 = 0.9$). The results of the \org\ morphology classification described in Section~\ref{sec:tag} are shown, for a slice through the simulation box, in Figure~\ref{fig:rhovsm}. The middle panel shows, in Lagrangian coordinates, the redshift-zero \org\ morphology indices $M$ of $256^2$ particles that inhabit a flat plane with equal $z$ in the initial lattice.  Each pixel of the square image corresponds to a particle.  The bottom panel shows the particles in Eulerian coordinates, again colored according to $M$.

\begin{figure}[htb]
  \begin{center}
    \includegraphics[width=\hsize]{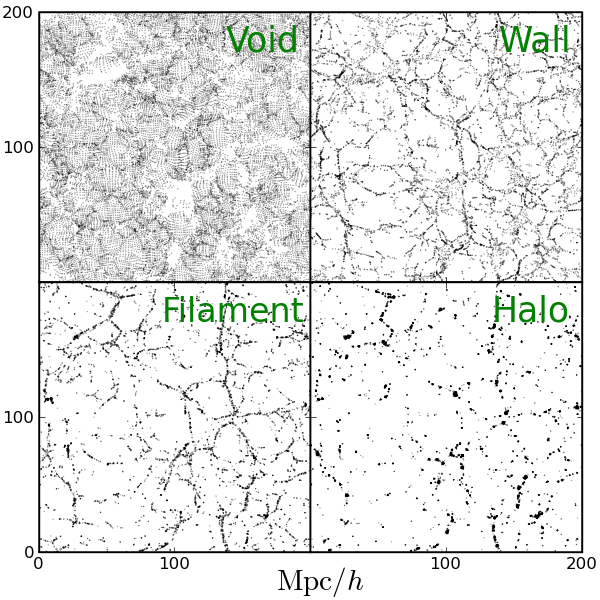} %.55
  \end{center}  
  \caption{The four components of the bottom panel of Figure~\ref{fig:rhovsm}, separated into different panels.}
  \label{fig:fourpanes}
\end{figure}

The top panel of Figure~\ref{fig:rhovsm}, plotted in Lagrangian space, shows
the VTFE density (described in Section~\ref{sec:group}) at each
particle.  The color scale is logarithmic, and a faint red contour is
added at $1+\delta_{\rm VTFE}=61$.  This density, perhaps surprisingly
low, divides halo from non-halo particles if one (wrongly) assumes
that ``haloness'' depends only on density.  More precisely, the
fraction of particles exceeding this density equals the fraction of
particles that are \org-identified halo particles.
Figure~\ref{fig:fourpanes} shows the four morphological components of the bottom panel of Figure~\ref{fig:rhovsm}, separated out for clarity.

In these figures, the \org\ identification of particle morphologies
accords with expectation, most obviously for void and halo particles.
For wall and filament particles, the situation is harder to assess in
a two-dimensional image, but again the classification looks reasonable.
A couple of small regions appear void-like in this two-dimensional
projection and yet are classified as walls; the shell-crossings
producing these walls happen to lie in the plane of the figure.  A
comparison of \org\ morphology to other morphology measures will be
the subject of a future study.

There is an interesting duality between the structures in the middle
and bottom panels of Figure~\ref{fig:rhovsm}.  This duality was noticed by investigators of the adhesion model of structure formation~\citep[e.g.,][]{KofmanEtal1990,KofmanEtal1992}. Halos in Lagrangian
space are large bubbles that are qualitatively like voids in Eulerian
space, although Eulerian voids are more polyhedral, whereas Lagrangian
halos are generally rounder.  Dually, voids in Lagrangian space are small,
as are halos in Eulerian space.  The situation with filaments and
walls is harder to see, but in Figure~\ref{fig:rhovsm} many Eulerian
filaments look, in Lagrangian space, like walls dividing bubbles (halos).

\subsection{Comparison to Friends-Of-Friends}
\label{sec:fof}

As part of the ``Haloes Gone MAD'' comparison project~\citep{kne11}, \org\ has been shown to be in general agreement with most of the standard
halo-finders in use today. To look at the details of how \org\ works,
particularly in the limit of very small halos where methods tend to
disagree the most, we create \org\ and \fof\ catalogs for a
100~$h^{-1}$Mpc $N$-body simulation with similar cosmology parameters as above (except $n_s = 0.93$ and $\sigma_8 = 0.81$) and at two different mass
resolutions. The high-resolution simulation has $512^3$ particles and
will be referred to in what follows as the 512 simulation. The
low-resolution simulation has $256^3$ particles with initial
conditions down-sampled from the 512 simulation, such that the only
difference between the initial density fields of the two simulations
is the resolution, and it will be referred to as the 256
simulation.

The \fof\ halo-finding method is one of the most widely-used and well-understood halo finders. 
It works by connecting particles that are separated by a distance smaller than some linking length, which is a parameter that we set to the typical value of 0.2 times the mean inter-particle separation. Though popular, \fof\ is not without issues. As noted widely in the literature, this method of linking particles into halos causes some fraction of the halos to be ``bridged'' such that two density peaks connected by a filament of particles are grouped into the same halo~\citep[see, e.g.,][]{luk09}. Additionally, there is some doubt as to whether \fof\ captures the full amount of particles involved in the collapse of the halo~\citep{and11}. This issue of the definition of halo edges is partly what motivated~\citet{kne11} to prefer $v_{\rm max}$ as a stable quantity by which to compare halo catalogs, as it largely ignores the outer parts of halos. 

\begin{figure}[htb]
\centering
\includegraphics[angle=90,width=\hsize]{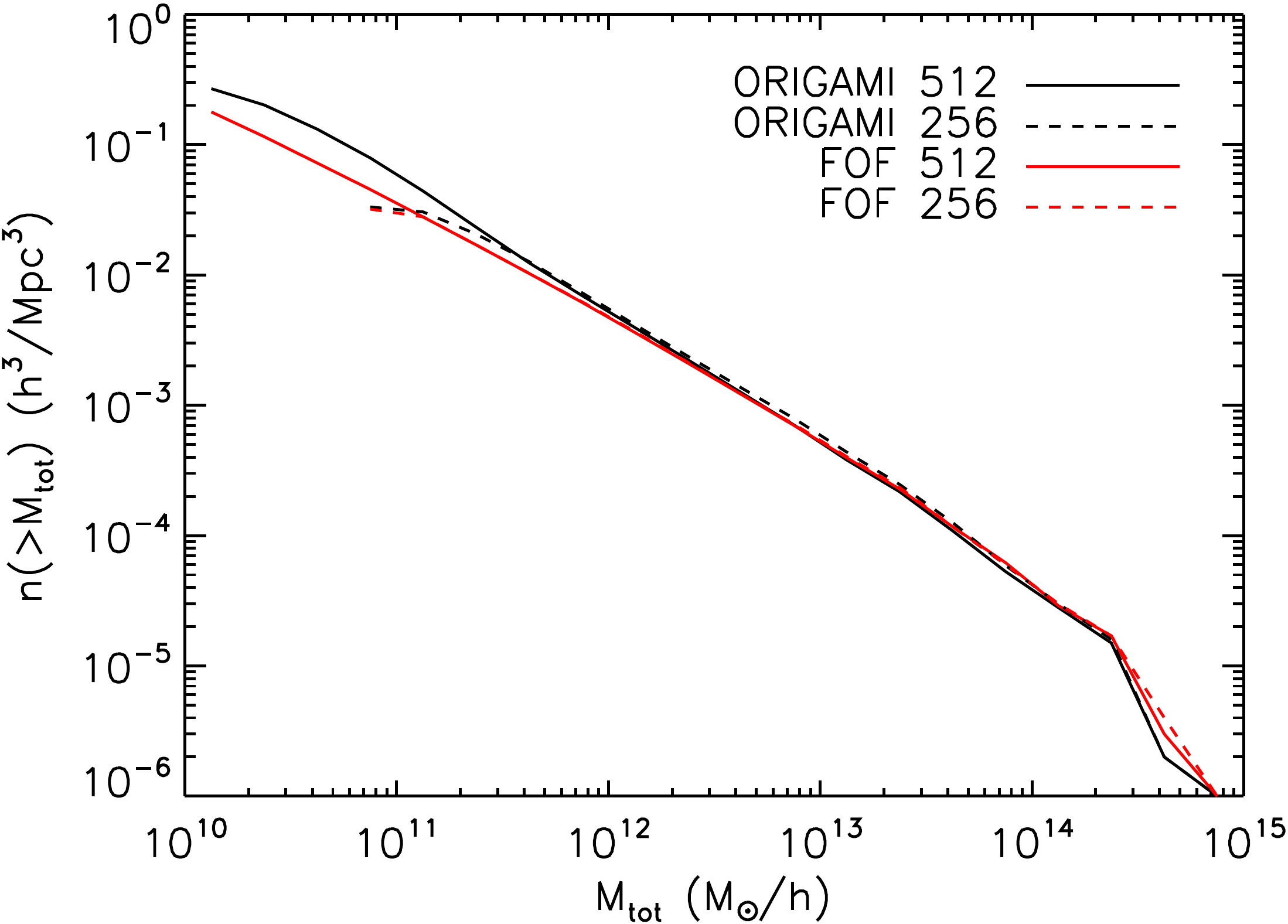}
\caption{Cumulative distribution functions of total mass for \org\ and \fof\ halos at both mass resolutions.\label{fig:mtot_all}}
\end{figure}

\begin{figure}[htb]
\centering
\includegraphics[angle=90,width=\hsize]{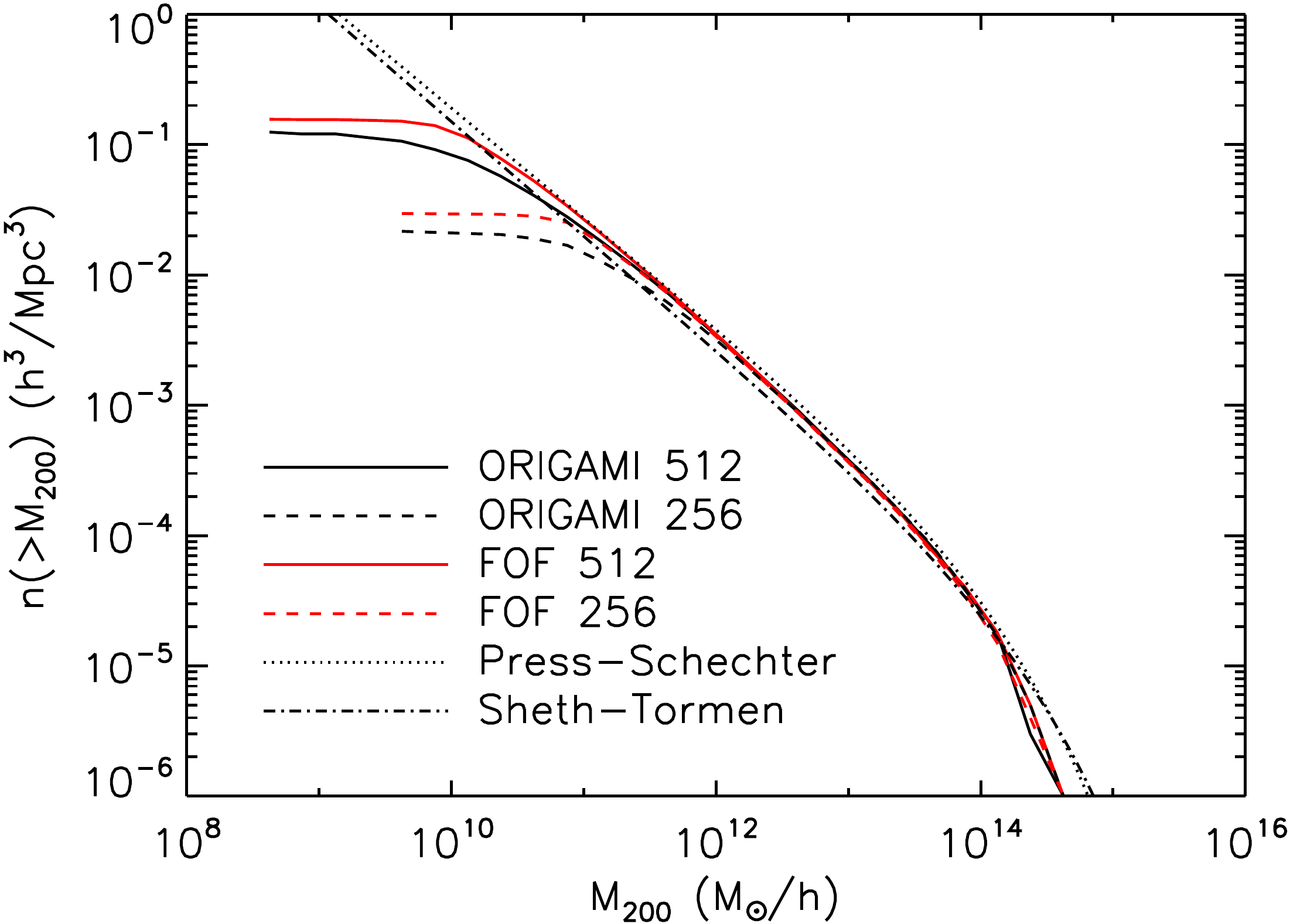}
\caption{Cumulative distribution functions of $M_{200}$ for \org\ and \fof\ halos at both mass resolutions. The theoretical curves of~\citet{	sheth99} and~\citet{pre74} are plotted for reference and are in good agreement.\label{fig:m200_all}}
\end{figure}

We first compare the cumulative mass functions. There is good agreement between \org\ and \fof\ in the total mass (Figure~\ref{fig:mtot_all}), with \org\ finding more very low mass halos (especially for the 512 simulation; see Section~\ref{sec:res}) and \fof\ finding slightly more very high mass halos. This situation changes when considering $M_{200}$ (Figure~\ref{fig:m200_all}); \org\ finds fewer low mass halos, as noticed in \citet{kne11}, when this definition of the halo mass is used. This is because there are many more \org\ halos that do not have a defined $M_{200}$, having no radius for which the density is greater than $200\rho_{crit}$ (as described in Section~\ref{sec:props}). This is largely due to \org\ halos being in general more diffuse and non-spherical than \fof\ halos and therefore more sensitive to the location of the halo center. As seen in Table~\ref{tab:nhalo}, \org\ starts with more halos than \fof\ at both resolutions -- and even more considering halos with fewer than 20 particles, though we do not consider them here -- but \org\ loses a much higher percentage of its halos when counting only those that have a defined $M_{200}$.

\begin{deluxetable}{cccc}
\tablewidth{0pt}
\tablecaption{Number of halos $N_h$ for different halo-finders and simulation resolutions\label{tab:nhalo}}
\tablehead{
\colhead{Halo-finder} & \colhead{Sim.} & \colhead{$N_h$ ($\geq 20$ particles)} & \colhead{$N_h$ (defined $M_{200}$)}}
\startdata
ORIGAMI & 256 & 33266 & 21632 \\
FOF & 256 & 32052 & 29612 \\
ORIGAMI & 512 & 269192 & 124870 \\
FOF & 512 & 178397 & 156686
\enddata %\tablecomments{}
\end{deluxetable}

\begin{figure}[htb]
\centering
\includegraphics[angle=90,width=\hsize]{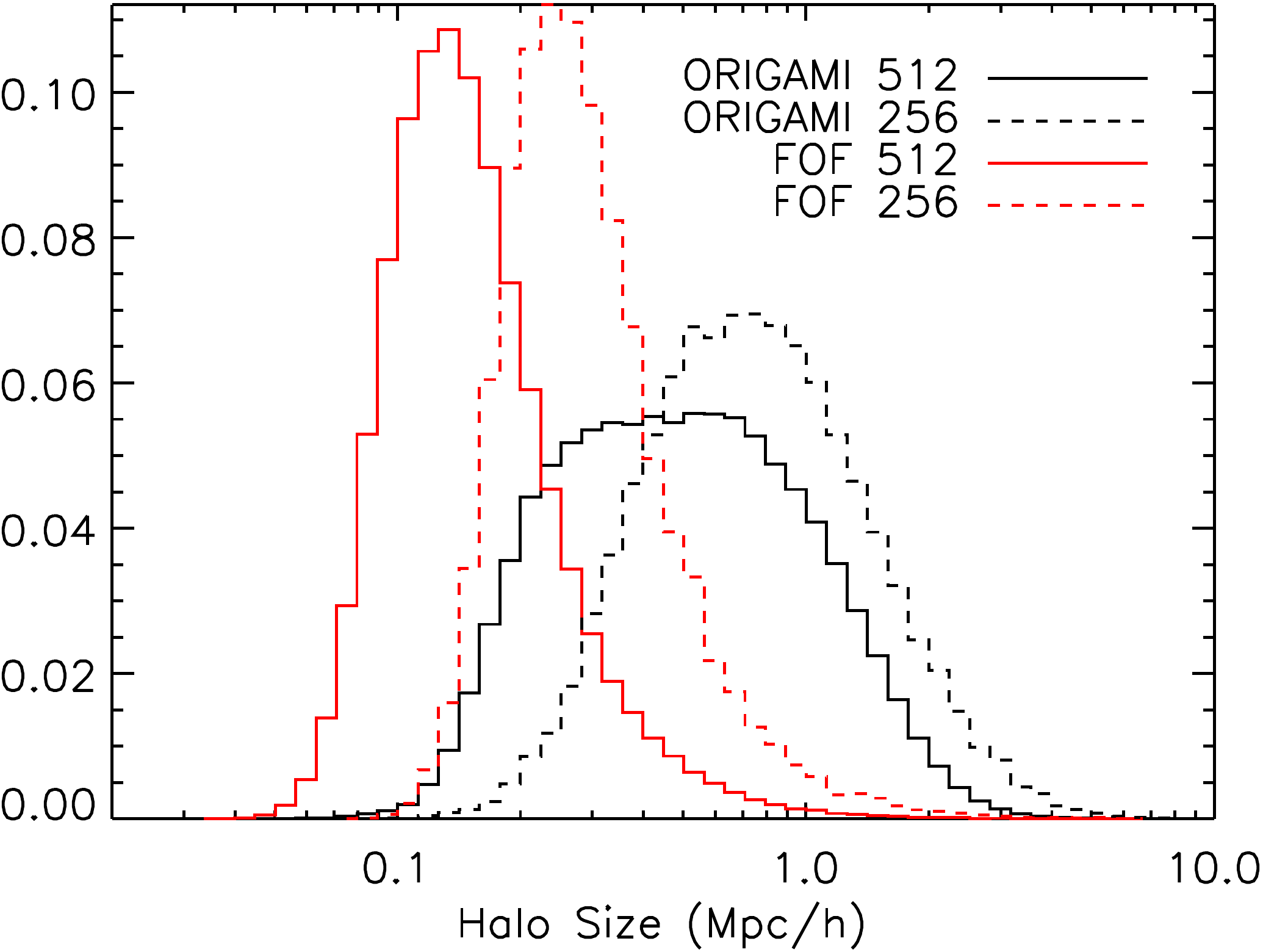}
\caption{Distribution of halo sizes (defined as the maximal diameter among Cartesian directions) for \org\ and \fof\ halos at both mass resolutions. \label{fig:size_all}}
\end{figure}

We now turn to the distributions of halo size, defined as the maximum Cartesian diameter in Section~\ref{sec:props}, and find that \org\ and \fof\ have very different distributions (Figure~\ref{fig:size_all}). For the 256 simulation, \fof\ halos have a distribution around 0.3~$h^{-1}$Mpc while \org\ halos have a slightly wider distribution with a mean of around 0.7~$h^{-1}$Mpc. Since the total mass functions are similar, it appears that \org\ halos are more diffuse and extend beyond the outer boundaries of \fof\ halos (which may be missing these edge particles~\citep{and11}), while at the same time \org\ halos sometimes (though not often) contain within them non-halo particles tagged with $M=2$ (filament) instead of $M=3$ (halo; see Section~\ref{sec:tag}). Recall that no unbinding procedure has been performed on either halo catalog, which would potentially remove the unbound particles from \fof\ halos, similar to how the interloping $M=2$ particles are ignored in the \org\ method.

\begin{figure}[htb]
\centering
\includegraphics[angle=90,width=\hsize]{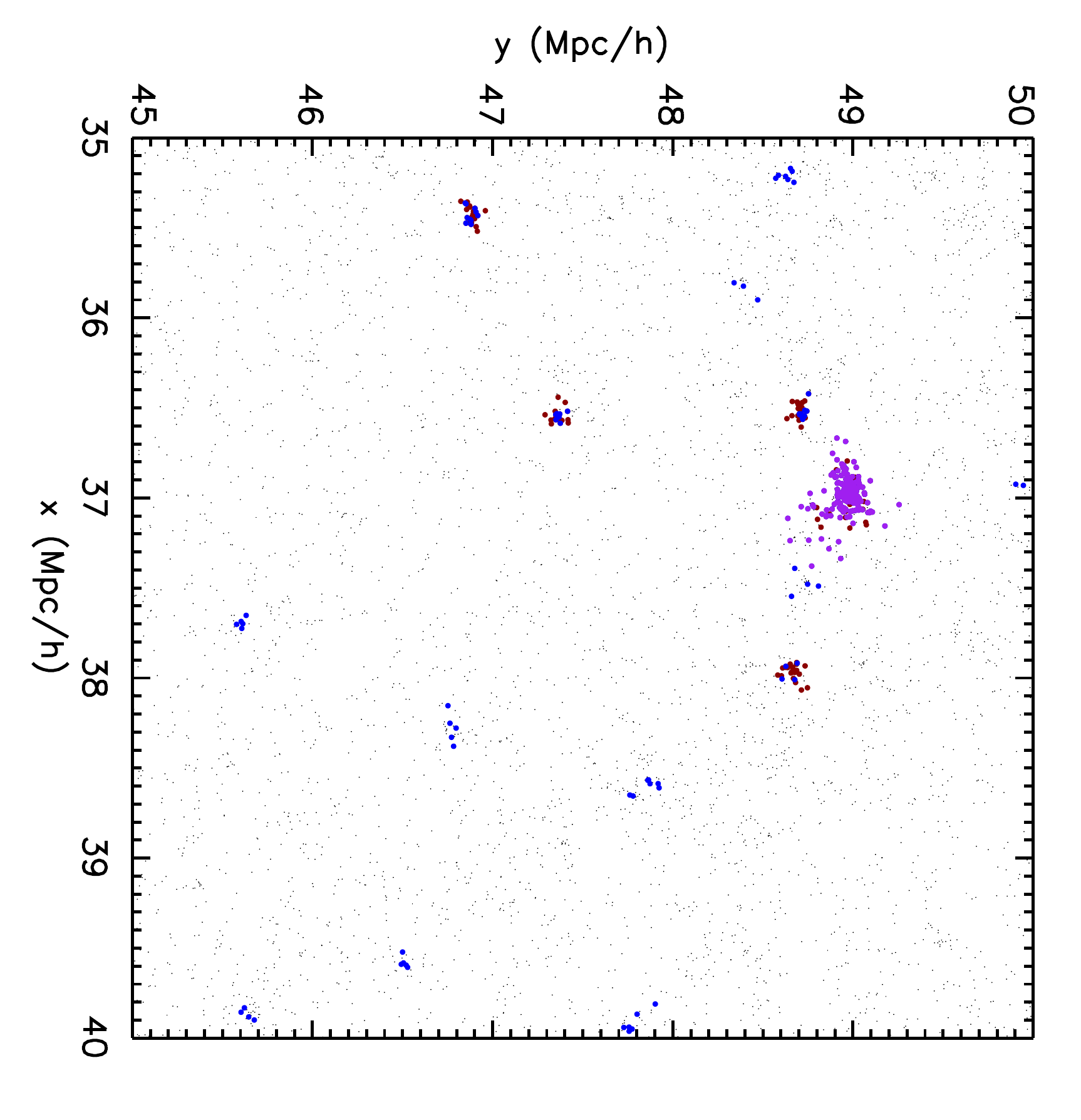}
\caption{Particle locations in a $5\times 5$ $h^{-1}$Mpc region of the 256 simulation, 40 $h^{-1}$Mpc deep in the $z$-direction. Over-plotted as larger dots are \fof\ halo particles in \emph{red}, $M=3$ halo-classified particles in \emph{blue}, and lastly \org\ halo particles (with at least 20 particles per halo) in \emph{purple}. The small \fof\ halos contain some, but fewer than 20, $M=3$ particles, and so do not become \org\ halos. \label{fig:zoom256}}
\end{figure}

The other effect causing the difference in halo sizes is that the very small \fof\ halos often contain a mix of $M=2$ and $M=3$ particles, such that fewer than 20 $M=3$ particles are grouped and therefore the group does not make it into the \org\ halo catalog. This can be seen in Fig.~\ref{fig:zoom256}, where a small region of the simulation containing one \org\ halo and a few small \fof\ halos is plotted showing \fof\ halo particles in red, \org\ halo particles (in which halos have at least 20 particles) in purple, and \org-classified $M=3$ particles in green.  Though some $M=2$ particles are close enough to their $M=3$ neighbors to be grouped by the \fof\ method, they do not meet the \org\ definition of a halo particle. 

\begin{figure}[htb]
\centering
\includegraphics[angle=90,width=\hsize]{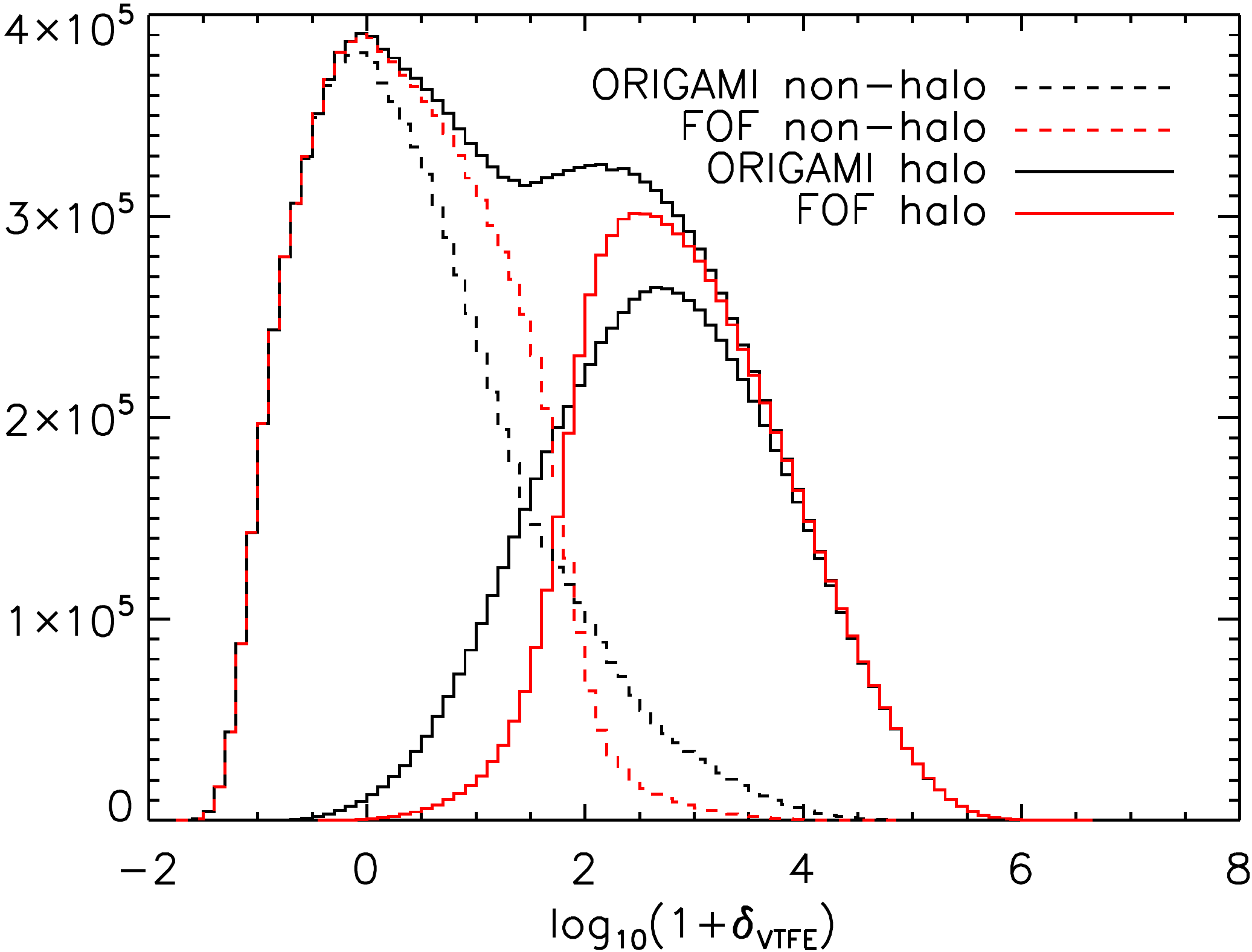}
\caption{Distribution functions of the VTFE particle density for \fof\ and \org\ halo and non-halo particles from the 256 simulation. \label{fig:rho256}}
\end{figure}

The VTFE density distributions for \fof\ and \org\ particles are shown in Figure~\ref{fig:rho256} for the 256 simulation. There is a more abrupt distinction between halo and non-halo particles for the \fof\ distributions than there is for the \org\ distributions, which accords with the common conception that the \fof\ linking length parameter corresponds to a density~\citep[but for a detailed study of this issue see][]{mor11}. It is interesting here to note that the double-peaked histogram of $\log(1+\delta_{\rm VTFE})$ is a particular feature of the mass-weighted nature of the density calculated from the Voronoi tessellation. As mass resolution increases and more small-scale features are resolved in a simulation, this peak grows (see, e.g., Fig.~\ref{fig:rho512} in Section~\ref{sec:res}), whereas an Eulerian (volume-weighted) measure of the density would wash out these features.

\begin{figure}[htb]
\centering
\includegraphics[angle=90,width=\hsize]{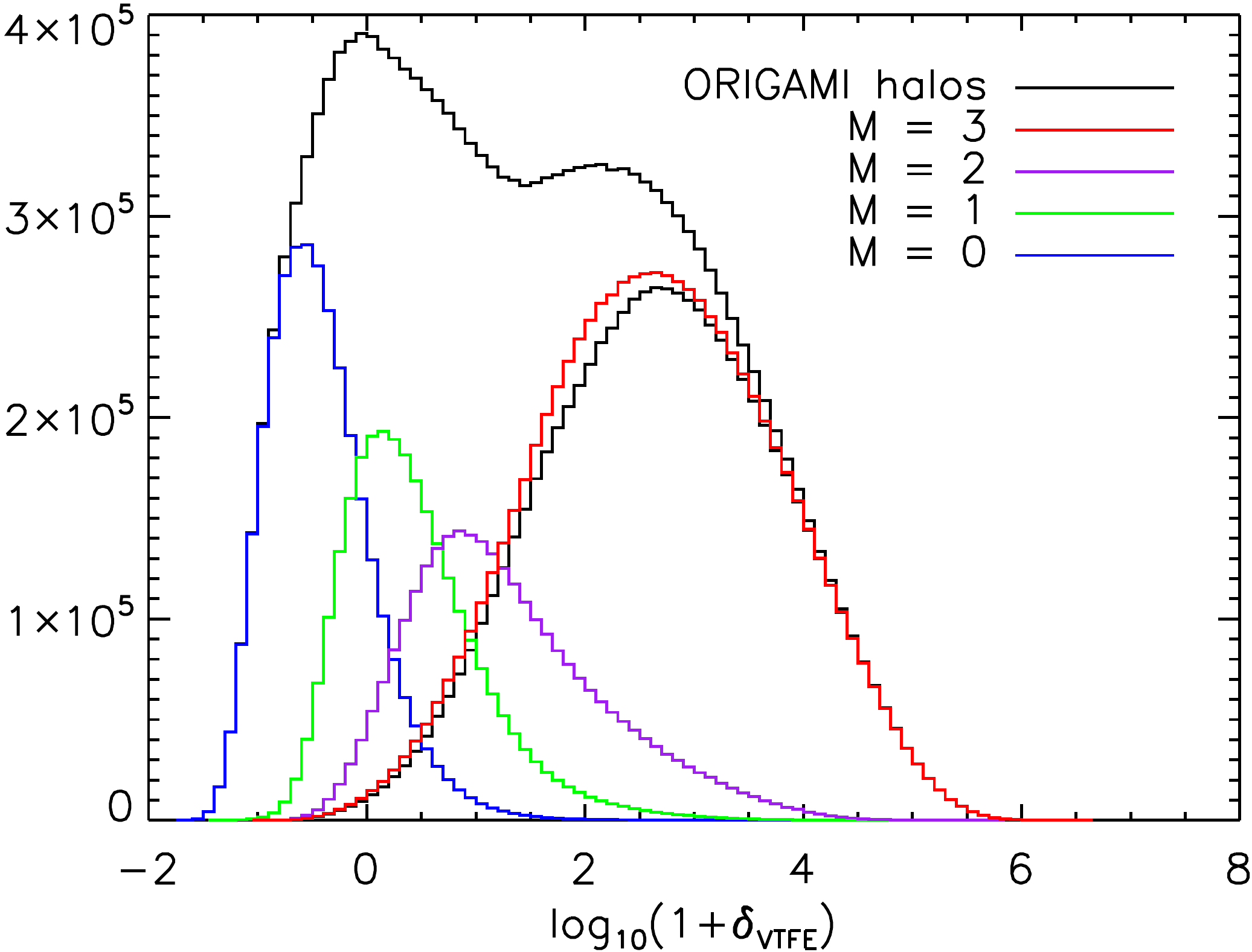}
\caption{Distribution functions of the VTFE density for particles tagged as $M=0$ (void, blue), $M=1$ (wall, green), $M=2$ (filament, purple), and $M=3$ (halo, red), for the $N=256$ simulation. The distribution function for \org\ halos is different than that for the $M=3$ particles because of the requirement that halos contain at least 20 particles. \label{fig:rhotag256}}
\end{figure}

In Figure~\ref{fig:rhotag256} the VTFE density distributions are plotted for \org\ particles separately based on morphology tag. As expected, the morphology tag is correlated with density, though there is much overlap, and in particular the $M=2$ particles have a long tail out to high densities. This tail likely corresponds to $M=2$ particles clustered near $M=3$ particles that become (small) \fof\ halos but contain too few $M=3$ particles to become \org\ halos (see Fig.~\ref{fig:zoom256}).

\subsection{Resolution Effects}
\label{sec:res}

\begin{figure}
  \begin{center}
    \includegraphics[width=\hsize]{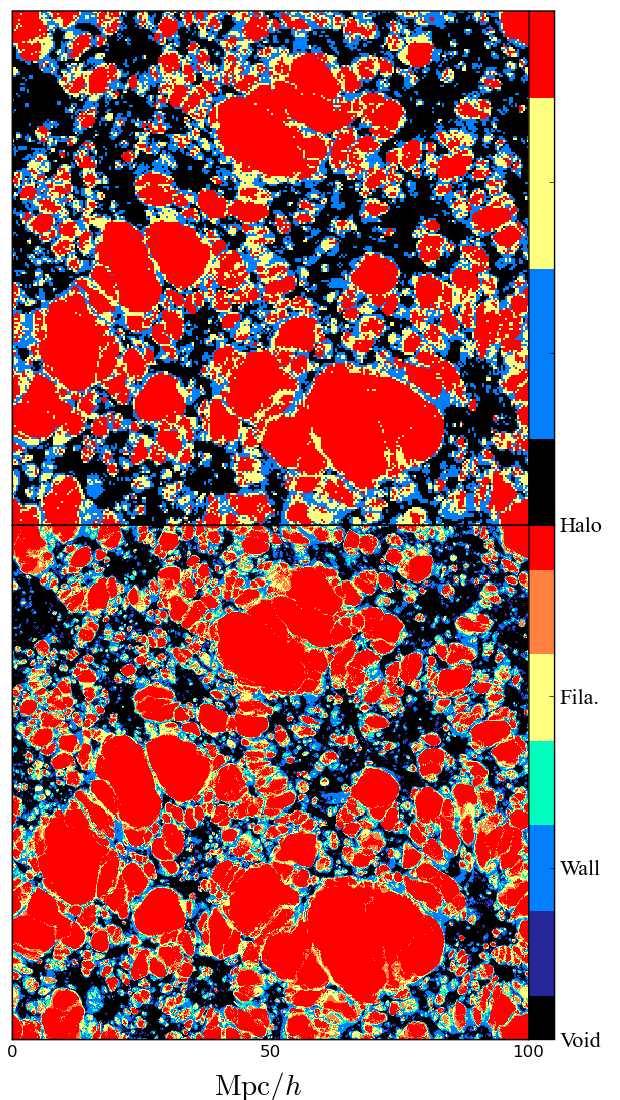} %.5
  \end{center}  
  \caption{The \org\ morphologies $M$, shown in Lagrangian
    coordinates, of initially two-dimensional sheets of particles in
    the 256 \emph{(top)} and 512 \emph{(bottom)} simulations.  In the top panel, each of the 256$^2$ pixels corresponds to a particle.  In the bottom    panel, the 512$^2$ pixels are colored according to the average
    \org\ morphologies of the two 512$^2$ sheets that comprise the
    256$^2$ sheet in the top panel.  Half-integers occur in the bottom
    panel because of this averaging.  For clarity, we show a slice
    with a relatively small fraction of halo particles.  }
  \label{fig:morph100}
\end{figure}

In general, we expect that as resolution increases, the boundaries of
large halos remain fairly constant and new, smaller halos with lower
masses are identified. Indeed, this is what \org\ finds, as shown in
Figure~\ref{fig:morph100}, which shows the \org\ morphologies of the same
redshift-zero Lagrangian sheet in the 256 and the 512 simulations.
Recall that the 256 simulation is the same as the 512 simulation, but with
coarsened resolution on the initial grid.  New halo regions appear in
the 512 simulation, particularly in void and wall regions of the 256
simulation.

However, the effect of resolution does not appear to be quite the same for the \org\ and \fof\ halo catalogs. As noted in Section~\ref{sec:fof}, the difference between the \org\ and \fof\ cumulative total mass functions is much greater for the 512 simulation than for the 256 simulation (see Figure~\ref{fig:mtot_all}), with \org\ finding noticeably more low mass halos. This means that the effect of increasing the simulation resolution adds many more \org\ halos than \fof\ halos. This effect is not seen in the $M_{200}$ mass function (Figure~\ref{fig:m200_all}) because these smaller halos continue to have an undefined $M_{200}$, and so the effect of increased resolution for both catalogs is the extension of the $M_{200}$ mass function down to lower masses.

\begin{figure}[htb]
\centering
\includegraphics[angle=90,width=\hsize]{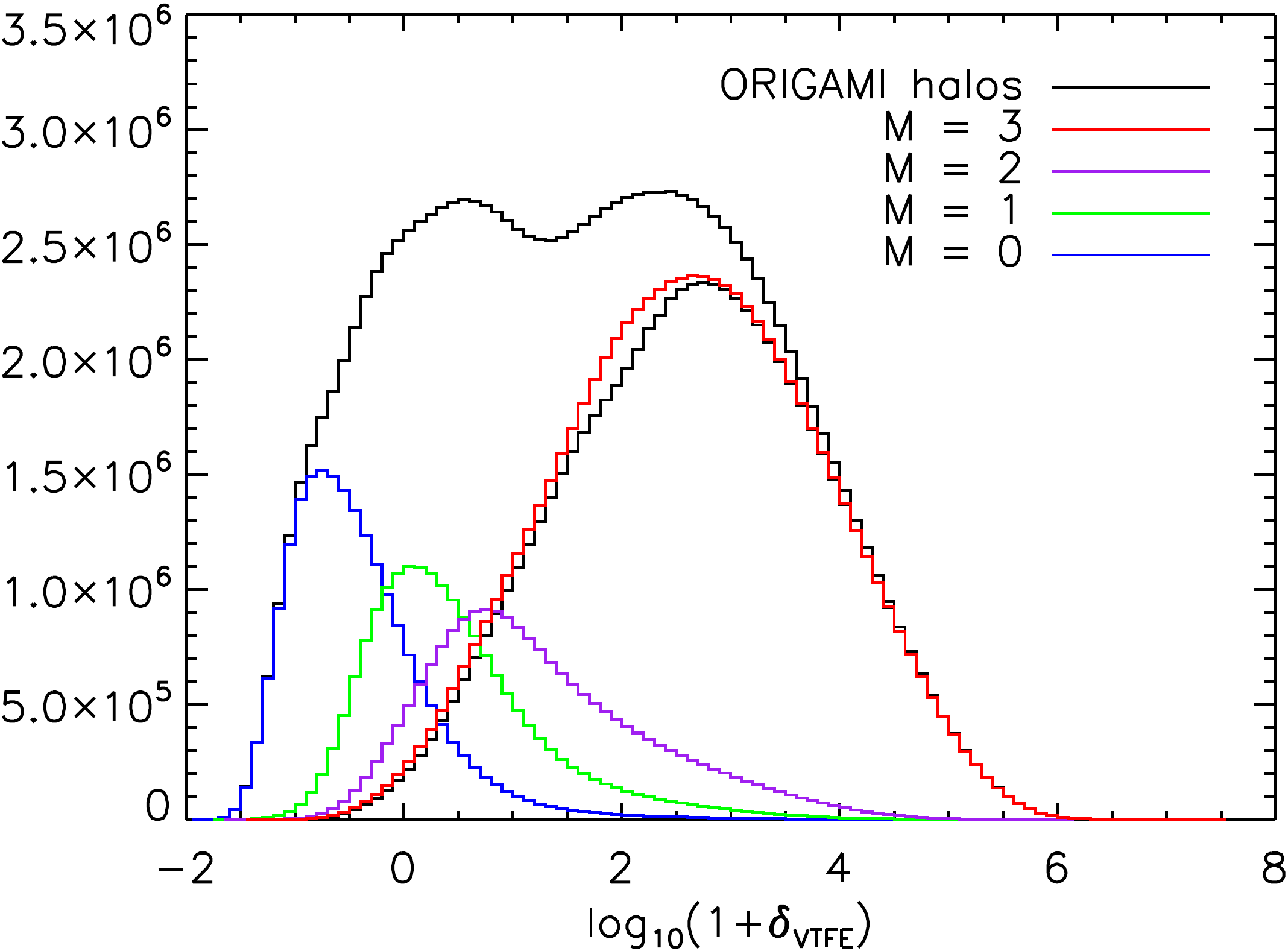}
\caption{Distribution functions of the VTFE density for particles tagged as $M=0$ (void, \emph{blue}), $M=1$ (wall, \emph{green}), $M=2$ (filament, \emph{purple}), and $M=3$ (halo, \emph{red}), for the $N=512$ simulation. \label{fig:rhotag512}}
\end{figure}

\begin{figure}[htb]
\centering
\includegraphics[angle=90,width=\hsize]{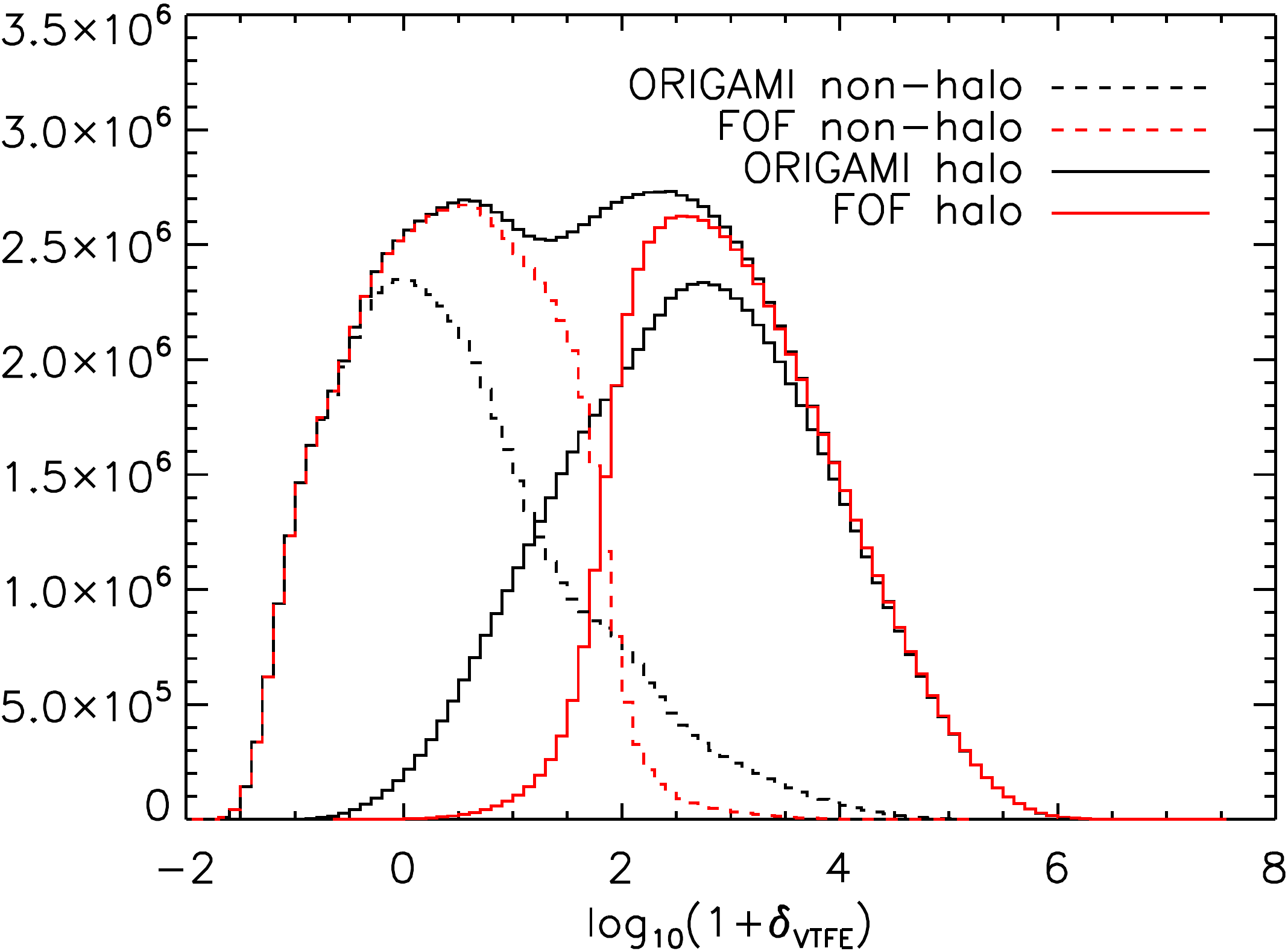}
\caption{Distribution functions of the VTFE particle density for \fof\ and \org\ halo and non-halo particles from the 512 simulation. \label{fig:rho512}}
\end{figure}

The reason why there are many more low mass halos in the 512 simulation is because a greater percentage of particles (56.6\%) are tagged as $M=3$ halo particles, compared to 47.7\% in the 256 simulation. This is partially related to the greater percentage of high density particles in the 512 simulation in general, as shown in Figure~\ref{fig:rhotag512}, compared to the 256 simulation in Figure~\ref{fig:rhotag256}. This increase in density also affects the \fof\ halo catalog, however it is to a much lesser degree: 47.3\% of particles are in \fof\ halos in the 512 simulation, compared to 42.7\% in the 256 simulation (see Figure~\ref{fig:rho512}).

If we look again at the distribution of halo sizes in Figure~\ref{fig:size_all}, we see that not only do both \org\ and \fof\ size distributions shift toward smaller halos, but additionally, the \org\ distribution becomes wider. We interpret this widening of the \org\ size distribution as being linked to the relative increase of low mass halos in the 512 simulation compared to \fof\ halos (Figure~\ref{fig:mtot_all}). The very small groups which had a mix of $M=2$ and $M=3$ particles in the 256 simulation, such that fewer than 20 $M=3$ particles were grouped, have a greater percentage of $M=3$ particles in the 512 simulation, so more of these small groups are counted as full halos containing at least 20 particles.

As a final note regarding resolution, we mention that in \org, the
resolution with which structures are found need not correspond to the
initial Lagrangian particle spacing, but could be a multiple of it.
This could be useful in building a sort of hierarchical morphology
tree from a single high-resolution simulation.  Also, if there is much
initial small-scale power in a high-resolution simulation, producing
undesirably many small \org\ halos, it could be useful to increase
the Lagrangian resolution used for \org\ morphology-tagging.

\section{DISCUSSION}
\label{sec:disc}

The current version of the algorithm requires a Delaunay tessellation to group the halo particles into constituent halos, but we would like to stress that \org\ is fundamentally a particle morphology tagger. The tagged halo (and filament, wall, and void) particles can be grouped in different ways; the tessellation method merely was expedient and sufficient for this paper. A Lagrangian-space method was also tried, which grouped halo particles that are connected on the initial Lagrangian grid as opposed to the final-conditions tessellation, but it was found that too many halo particles are connected on the Lagrangian grid which later form distinct structures. One possibility to improve the particle-grouping step would be to add bookkeeping that keeps track of \emph{which} other particles a given particle has crossed paths with; though it is not necessarily true that these particles all end up in the same bound structure, the information could be useful.

Though a comparison of the \org\ morphology classification to other methods is left to future work, we are encouraged by the results so far: by eye, it looks as if \org\ does very well in identifying filament, wall, and void particles as well as halos. The task of grouping these particles into individual filaments, walls, and voids, however, will likely prove more challenging than for halos. It may be worthwhile to use \org\ in concert with another morphology identification method that can be modified to take advantage of the \org\ particle classification.

\org\ is successfully able to calculate halo catalogs for cosmological dark matter $N$-body simulations, however, there may be some limits to its applicability. Due to its nature, \org\ is unable to find groups in observations because it relies on information about the initial state of the system. Similarly, it would be difficult, though possible in principle, to apply \org\ to simulations with baryons or with irregular (such as `glass') initial conditions because the particles aren't initially aligned on a regular grid. However, note that the idea of using caustics to identify the outer regions of groups has already been applied to measuring the masses of galaxy clusters~\citep{dia99,dia05}. Finally, \org\ is currently unable to distinguish subhalo particles from their parent halo, though it is possible that using velocity information will allow \org\ to identify substructure.

As mentioned in the introduction, the mathematical field of paper origami has developed quite recently~\citep{Lang1996,Hull1994,Hull2002,Hull2006}.  In so-called flat origami, the final product is constrained to lie in a plane after folding, likely with some regions in which many layers of paper overlap.  A set of creases in the initial paper is only foldable into a flat origami design if the creases obey a set of laws.

Some of these laws are irrelevant to large-scale structure because of the different dimensionality and the inhomogeneous stretching that occurs, but a law that may be relevant is two-colorability. In two-dimensional flat origami, polygons in the paper must end up facing either up or down, so the tessellation formed by the initial crease pattern must be two-colorable, one color painting the ``up'' polygons, and the second color painting the ``down.''
In analogy, Lagrangian space can be tessellated with three-dimensional
regions bordered by caustics.  In this case, we speculate that regions could
be successfully colored according to their two possible orientations or
chiralities, i.e.\ according to whether the three initially right-hand-oriented
axes are left- or right-hand oriented. This issue will be the subject of a future study: two-colorability would be quite special for a three-dimensional tessellation, for which arbitrarily many colors (not only four, as in two dimensions) can in principle be required~\citep[e.g.,][]{Wilson2002}.

\section{CONCLUSION}
\label{sec:conc}

The \org\ structure-finding algorithm identifies particles as belonging to a halo, filament, wall, or void by determining whether they have crossed paths with their Lagrangian neighbors along 3, 2, 1, or 0 orthogonal dimensions, respectively.
In the present implementation, halo-classified particles that are connected on a Delaunay tessellation are grouped into individual halos, though in principle there are many ways in which halo particles may be grouped. Long strings or structures of over-connected halo particles are prevented by first requiring that halos contain at most one halo ``core,'' or set of connected halo particles, that are above some density threshold. 

We have compared \org\ halo catalogs to \fof\ catalogs at two different mass resolutions and found that the mass functions agree very well; more comparisons to other halo-finding algorithms are given in~\citet{kne11}. Though the mass functions largely agree, \org\ halos are in general much larger than \fof\ halos, suggesting that \org\ halos are a bit more diffuse and spread out. Additionally, the smallest \fof\ halos contain many particles that \org\ classifies as belonging to a filament, resulting in \org\ finding fewer small halos. This result stresses the difference between a density-based definition of a halo and the \org\ method of detecting the occurrence of shell-crossing.

The effect of increasing the simulation resolution is in general as expected, with more structures popping up at smaller scales. Both the \org\ and \fof\ methods find a larger fraction of halo particles at a higher mass resolution, but this effect is greater for \org. For the  mass function, this results in \org\ finding an increased fraction of low mass halos compared to \fof. Additionally, the \org\ distribution of halo sizes gets wider as resolution increases, while the \fof\ distribution stays roughly the same shape, and both shift to smaller sizes. This resolution effect is not seen in the $M_{200}$ mass function, suggesting that it largely occurs at the outer edges of the halos and for low-density halos, both of which have little effect on $M_{200}$.

With \org, we have defined the boundary of a halo as being located at the outer phase-space caustic. This has produced some interesting differences in the sizes and general characteristics of \org\ and \fof\ halos, which diagnostics that depend only on the central cores of halos tend to miss. In other algorithms, the definition of a halo edge usually depends strongly on a free parameter, but in \org\ the definition of halo particles is parameter-free. Instead of depending solely on density, \org\ finds halos by looking for folds in phase-space.

\acknowledgments

B.L.F. and M.C.N. thank Alex Knebe, Steffen Knollmann, Gustavo Yepes, and Justin Read for hosting the ``Haloes going MAD'' workshop at the beautiful La Cristalera de Universidad Autonoma de Madrid. We are grateful to Miguel Aragon-Calvo for providing the simulations and much help and feedback while this algorithm was being developed. The authors thank Peter Behroozi, Marcel Haas, David Larson, Guilhem Lavaux, Nuala McCullagh, and Istvan Szapudi for helpful discussions and comments. M.C.N. thanks Robert Lang for an inspiring colloquium about paper origami and further discussions. The authors acknowledge financial support from the Gordon and Betty Moore Foundation and the NSF OIA grant CDI-1124403.

\bibliographystyle{hapj}
\bibliography{refs}

\end{document}